# Interlayer Five-Spin Polaron in Superconducting Bilayer Nickelates

Jiarui Li[1†], Christopher T. Parzyck[1], Eder G. Lomeli[1], Yidi Liu[1,2], Taehun Kim[3], Heemin Lee[4], Zengqing Zhuo[5], Eun Kyo Ko[1,6], Yaoju Tarn[1,6], Cheng-Tai Kuo[4], Ronny Sutarto[7], Chunjing Jia[8], Vivek Thampy[4], Jonathan Pelliciari[3], Wanli Yang[5], Brian Moritz[1], Yijun Yu[1,6*], Jun-Sik Lee[4], Valentina Bisogni[3], Thomas P. Devereaux[1,9†], Harold Y. Hwang[1,6†], Wei-Sheng Lee[1†]

[1]Stanford Institute for Materials and Energy Sciences, SLAC National Accelerator Laboratory, Menlo Park, CA 94025, USA

[2]Department of Physics, Stanford University, Stanford, CA 94305, USA

[3]National Synchrotron Light Source II, Brookhaven National Laboratory, Upton, NY, 11973, USA

[4]Stanford Synchrotron Radiation Lightsource, SLAC National Accelerator Laboratory, Menlo Park, CA 94025, USA

[5]Advanced Light Source, Lawrence Berkeley National Laboratory, Berkeley, CA 94720, USA

[6]Department of Applied Physics, Stanford University, Stanford, CA 94305, USA

[7]Canadian Light Source, Saskatoon, Saskatchewan S7N 2V3, Canada

[8]Department of Physics, University of Florida, Gainesville, FL 32611, USA

[9]Department of Materials Science and Engineering, Stanford University, Stanford, CA 94305, USA

[†]jiaruili@stanford.edu, tpd@stanford.edu, hyhwang@stanford.edu, leews@stanford.edu

*Present address: State Key Laboratory of Surface Physics and Department of Physics, Fudan University, Shanghai, China

**Abstract**

The discovery of high-$T_c$ superconductivity in Ruddlesden-Popper nickelates has sparked substantial effort towards understanding unconventional electronic states beyond a traditional cuprate-like $d^9$ configurational ground state. An understanding of the interplay between magnetic ground states and multi-orbital physics is key for establishing a microscopic mechanism for superconductivity. In the bilayer nickelates, spin density wave (SDW) order is a prominent feature in the non-superconducting regime, yet its relation to superconducting pairing remains an open question. Here, we use resonant x-ray scattering to examine the existence of SDW order in superconducting bilayer nickelate thin films $La_2PrNi_2O_7$ (LPNO). Comparing superconducting and oxygen-deficient LPNO thin films, we find that superconductivity occurs in SDW-free, oxygen-stoichiometric regions, whereas oxygen-deficiency promotes SDW order, indicating phase segregation of SDW and superconductivity. Furthermore, Ni-$L_3$ and O-$K$ edge spectroscopy reveals distinct electronic structures – particularly along the *c*-axis – between the two phases. Our results identify oxygen stoichiometry as a key parameter controlling interlayer coupling and thus the electronic structure of bilayer nickelates. In concert with theory, we propose that a ligand hole primarily resides at the inter-bilayer apical oxygen, forming a robust interlayer five-spin polaron state, which serves as the ground state for superconducting bilayer nickelates.

**Main text**

The recent discovery of high-temperature superconductivity in several nickelates has expanded the materials family of unconventional superconductors[1–8]. In cuprate and infinite-layer nickelates close to the $3d^9$ electronic configuration, superconductivity coexists with short-range antiferromagnetic fluctuations[9–11]. A different picture emerges for the Ruddlesden-Popper (RP) phases of nickelates ($RE_{n+1}Ni_nO_{3n+1}$, where RE = rare earth). These materials present formally very different $d$ electron counts, with a nominal $d^{7+1/n}$ configuration. Long-range spin density wave (SDW) order has been widely reported in multiple compounds at ambient conditions[12–19]. This raises an intriguing question about the relationship between SDW order and superconductivity appearing at high pressure in $n$ = 2, 3 RP phase nickelates. Recent studies suggest a seemingly competitive relationship between density wave order and superconductivity – superconductivity emerges under high pressure, while SDW is suppressed[20–22]. However, spectroscopic study of the microscopic interplay between these phases is hindered by the high-pressure environments in the diamond anvil cells.

Recently, ambient pressure superconductivity above 40 K has been realized in thin films of bilayer nickelate $(La,Pr)_3Ni_2O_7$ under compressive strain[6,23–25]. However, bilayer nickelates are known to be chemically unstable towards apical oxygen vacancies that are detrimental to superconductivity[23,25–28]. A strongly oxidizing environment, applied either during growth or through post-growth annealing, is required to stabilize superconductivity by filling oxygen vacancies[23,25,26]. Post-growth annealing was also able to traverse metallic, superconducting, insulating phases via continuous tuning of oxygen stoichiometries[29]. By contrast, the SDW order has been found to be a robust phase in bilayer nickelate thin films across a large range of strain and oxygen stoichiometries, with a characteristic wavevector $\boldsymbol{Q}_{\mathrm{SDW}}$ = (0.25, 0.25) and a transition temperature around 150 K, comparable to bulk samples[14–16,19]. Therefore, thin film platforms enable a direct comparison between the superconducting and non-superconducting electronic ground states and the interplay between SDW order and superconductivity in the bilayer system under ambient conditions.

In this work, we investigate the interplay between SDW order and superconductivity through a comprehensive resonant x-ray scattering and spectroscopic approach near the Ni-$L_3$ and O-$K$ edges to reveal the electronic and magnetic ground states in multiple $La_2PrNi_2O_{7-x}$ (LPNO) thin films with different oxygen stoichiometries. We first identify an inhomogeneous distribution of SDW domains, confined to a few localized, sparse spots across the superconducting (SC) samples. Conversely, in a non-superconducting, oxygen-deficient (O-def) sample, the SDW is uniformly distributed and strong across the entire film, suggesting that the SDW is absent in oxygen-stoichiometric superconducting LPNO, and appears upon oxygen loss. Having established this context, we explore the electronic structure of the superconducting ground state via x-ray spectroscopy. We find that the superconducting phase resembles a metallic (small- or negative-charge-transfer) ground state with primarily Ni $d^8$ and significant oxygen ligand hole character. Moreover, in contrast to previous reports, we observe several distinct localized excitations at 0.4 eV and 1.5 eV only in oxygen deficient films, while they are absent in oxygen-stoichiometric, superconducting samples[30,31]. This behavior is consistent with enhanced localization and reduced charge screening upon oxygen loss. O-$K$ edge x-ray absorption spectroscopy (XAS) and resonant inelastic x-ray scattering (RIXS) further reveal that oxygen deficiency modulates the pre-edge spectral weight associated with both an in-plane Zhang-Rice-singlet-like state, as well as an apical ligand-hole state. Both Ni-$L_3$ and O-$K$ edge RIXS maps demonstrate that the $c$-axis electronic structure is highly sensitive to oxygen stoichiometry, underscoring its important role in determining the ground state. Our findings establish that SDW order is associated with local oxygen defects rather than the superconducting phase in bilayer nickelates. We propose that a robust interlayer Ni($d^8$)-O($L$)-Ni($d^8$) five-spin polaron state locks out-of-plane $3d_{z2}$ Ni charge and spin configurations, leaving in-plane $3d_{x2\text{-}y2}$ Ni orbitals near a half-filled regime, reminiscent of the $3d^9$ character in cuprate and infinite-layer nickelate systems.

**SDW and superconductivity phase segregation**

We first examine whether the SDW is intrinsic to superconducting LPNO grown on $SrLaAlO_4$ (SLAO) substrate (-2% compressive strain) using resonant x-ray scattering at Ni-$L_3$ resonance. In contrast to the widely reported SDW order observed in bulk and thin-film bilayer nickelates[14–16], we detect no SDW signature within our sensitivity at the putative SDW wavevector $\boldsymbol{Q}_{SDW}$ = (0.25, 0.25), at the center of the superconducting film (**Fig. 1a**). Instead, SDW peak could be observed at some sample positions (**Supplementary Fig. 1, 2**), with the same characteristic wavevector and peak shape reported previously[14–16]. Spatial mapping reveals that the SDW signal is localized in sparse, sub-millimeter patches, predominantly near the sample edges or beneath the gold electrodes (**Fig. 1b**, and **Supplementary Fig. 1**), while most areas of the film remain SDW-free.

Oxygen non-stoichiometry is a natural candidate to explain the material inhomogeneity. Previous studies in bilayer nickelates have reported a tendency toward oxygen loss[23,27], which could lead to an inhomogeneous SDW. We test this hypothesis with the same spatial mapping on a deliberately oxygen-deficient (O-def), non-superconducting LPNO film. In contrast to the superconducting sample, the O-def film exhibits a strong, uniform SDW response throughout the entire film (**Fig. 1c, d**), with a substantially higher average intensity. Because the x-ray beam footprint (~200 μm) averages over many microscopic regions, the pervasive signal indicates that SDW order occupies a large volume fraction in the oxygen-deficient films.

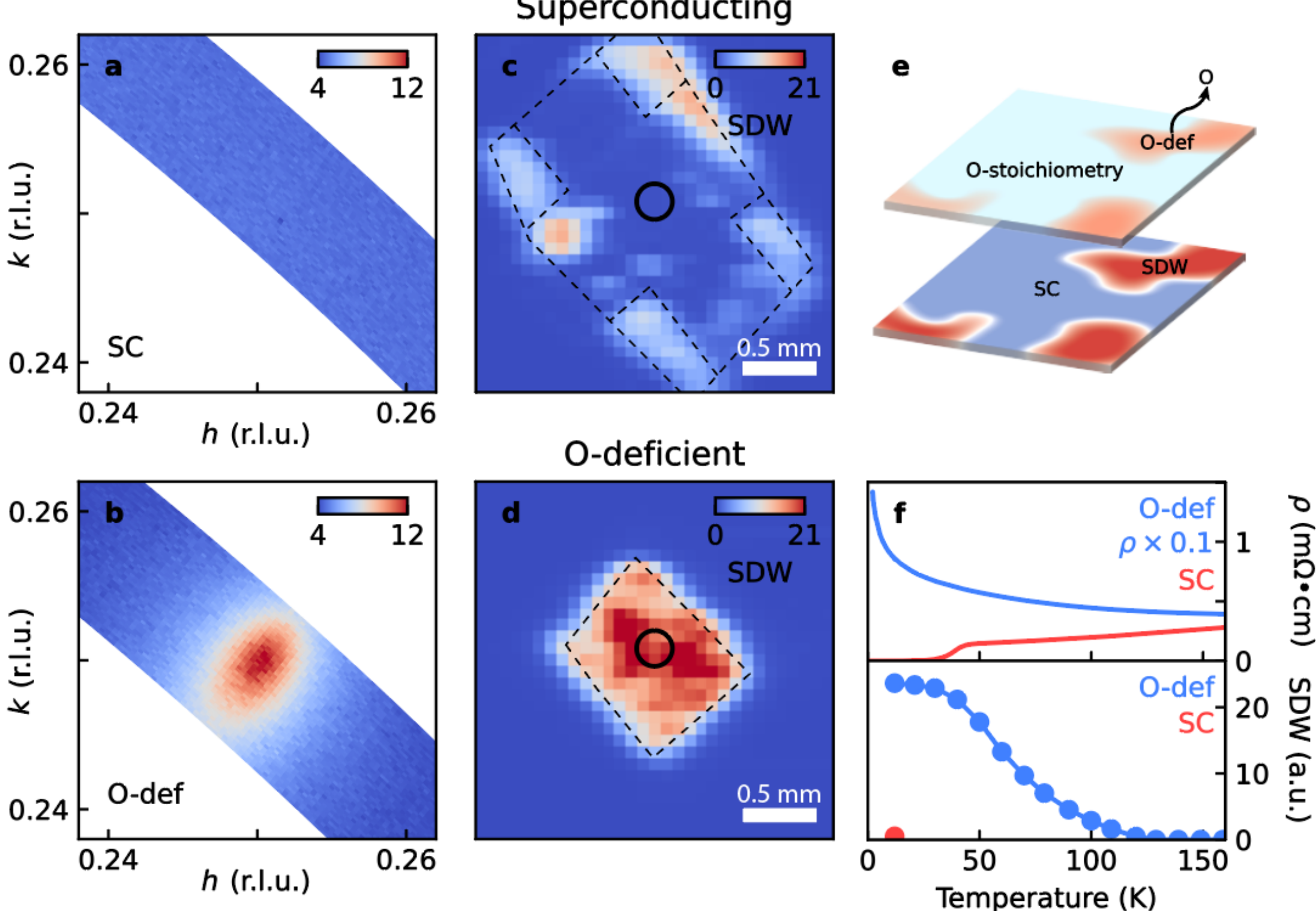

**Figure 1. Spatial heterogeneity of spin density wave (SDW) order and superconductivity in $La_2PrNi_2O_{7-x}$ thin films (LPNO). a**, Ni-$L_3$ resonant x-ray scattering intensity near $\boldsymbol{Q}_{SDW}$ = (0.25, 0.25) measured at the center on superconducting (SC, circle in **b**). **b**, Spatial map of the SDW peak intensity for the SC film showing highly inhomogeneous distribution of SDW domains. **c, d**, Resonant x-ray scattering intensity near $\boldsymbol{Q}_{SDW}$ = (0.25, 0.25) and the spatial map the SDW peak intensity for the oxygen-deficient, non-superconducting film (O-def). Dashed lines outline the film area. Both maps are taken at 11 K. Scale bar: 0.5 mm. Photos of the SC sample (**b**) and O-def sample (**d**) are shown in **Supplementary Fig. 1. e**, Schematic illustration of the oxygen-stoichiometry-driven phase segregation into oxygen-stoichiometric SC regions and O-def SDW regions. **f**, Temperature dependence of resistivity ($\rho$, top) and SDW intensity (bottom) for both samples. The resistivity of the O-def sample is scaled by 0.1. Center of SC sample remains SDW-free down to the base temperature.

These results point to oxygen-stoichiometry-driven phase segregation between superconductivity and SDW order as illustrated in **Fig. 1e**. Oxygen-stoichiometric regions are SDW-free and host superconductivity, whereas oxygen-deficient regions stabilize SDW order and remain non-superconducting. This interpretation is supported by several additional observations. First, the SDW peak profiles, transition temperatures, and temperature dependences within the SDW domains of the superconducting sample are found to be similar to those in the O-def sample (**Fig. 1f**, **Supplementary Fig. 2, 3a**). This is consistent with a common SDW phase originated from oxygen-deficient regions rather than a distinct SDW state. Second, the SDW peak intensity exhibits a smooth temperature dependence across the superconducting transition without discernible anomalies (**Supplementary Fig. 3b**). Third, histograms of the mapped SDW intensity (**Supplementary Fig. 3c**) show that oxygen-stoichiometric samples are dominated by near-zero SDW intensity with an exponential tail to finite intensity, whereas the O-def film exhibits a broad Gaussian-like distribution centered at large intensities, consistent with a high SDW volume fraction. The weak SDW domain in the oxygen-stoichiometric sample suggests sparse local oxygen deficiency that could be assisted by local defects. Finally, the phase segregation scenario is also supported by the granular superconductivity upon oxygen deficiency[29].

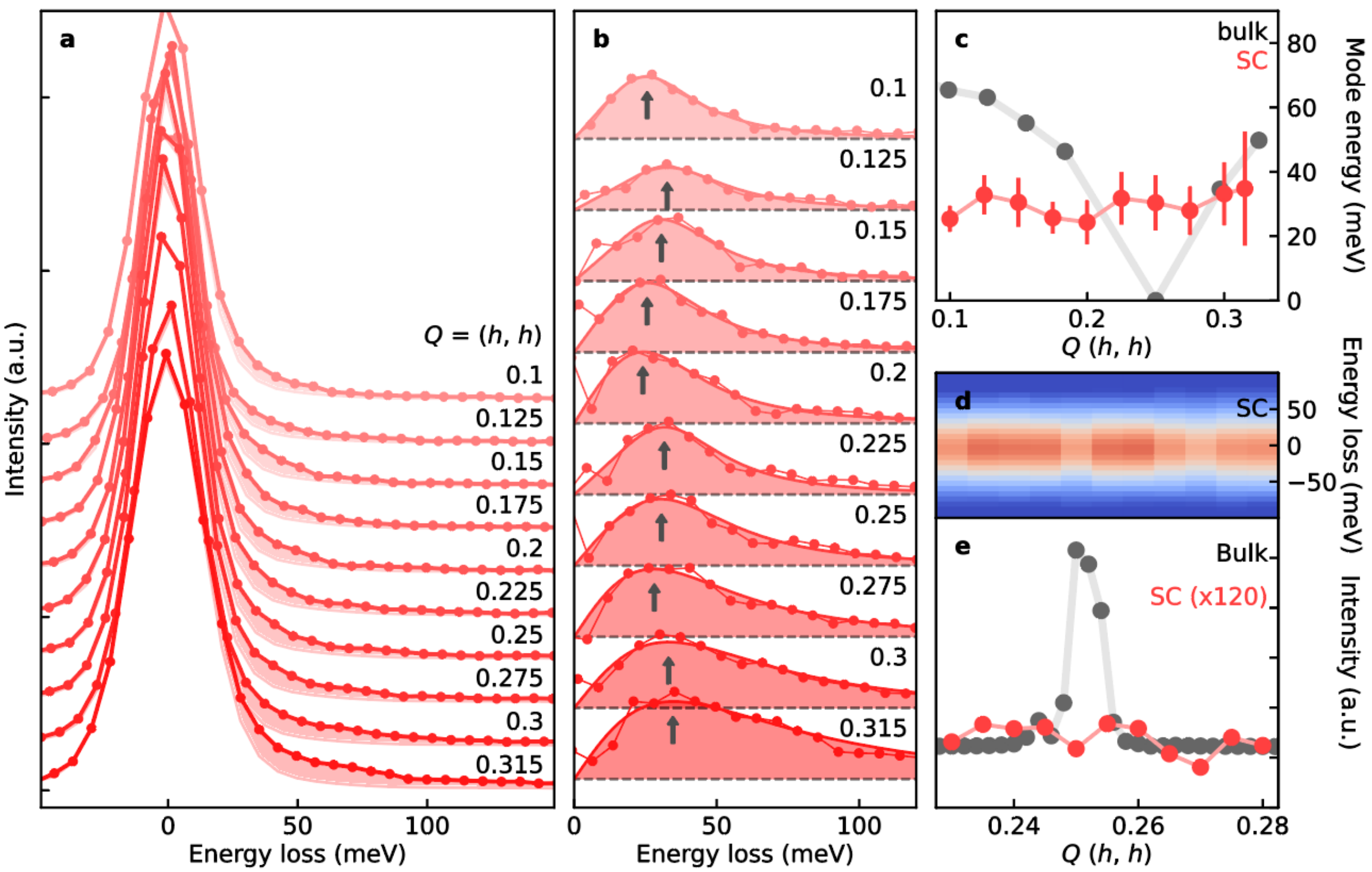

**Figure 2. Low energy excitations in superconducting $La_2PrNi_2O_7$. a,** Momentum-dependent RIXS spectra along the ($h$, $h$) direction from superconducting LPNO. The shaded region highlights the additional spectra weight beyond the tail of the elastic line. **b,** The elastic line subtracted spectra highlighting the low energy excitations. The shaded area represents the fit to a damped harmonic oscillator model. The arrows mark the maximum intensity of the fitted mode. **c,** Dispersion of the low energy mode in superconducting LPNO sample extracted from fitting the RIXS spectra. The magnon dispersion reported in bulk $La_3Ni_2O_{7-x}$ is also shown for comparison[14]. **d,** RIXS intensity map near the quasi-elastic region for superconducting samples. **e,** Momentum-dependent quasi-elastic intensity in the RIXS spectrum. The data for bulk $La_3Ni_2O_{7-x}$ is also shown for comparison[14].

Magnetic excitations serve as a direct fingerprint of a material's magnetic ground state, where their dispersion encodes the underlying order and quantifies the dominant exchange interactions. Previous studies have shown a dispersive magnon across the Brillouin zone rising out of the $\boldsymbol{Q}_{\mathrm{SDW}}$ = (0.25, 0.25) SDW order[30–32]. We probe the momentum-dependent low-energy excitations in both SC and O-def LPNO using resonant inelastic x-ray scattering (RIXS). **Figure 2a** shows the momentum dependent RIXS spectra acquired in the center of the superconducting sample along the ($h$, $h$) direction, which cuts through the expected $\boldsymbol{Q}_{\mathrm{SDW}}$. The spectra are dominated by the strong elastic line with additional spectral weight manifested as a tail at the foot of the elastic line. Upon subtraction of the elastic line (measured on a carbon tape, see also **Supplementary Fig. 4**), a broad excitation centered at ~30 meV is observed (**Fig. 2b**). This excitation shows no discernible dispersion as a function of momentum, suggesting its origin is distinct from the (0.25, 0.25) SDW (**Fig. 2c**). We also examine the momentum distribution of the quasi-elastic intensity in RIXS spectra, which provides higher sensitivity to SDW order than resonant elastic x-ray scattering by rejecting background arising from fluorescence and other inelastic signals. As shown in **Fig. 2d**, the lack of quasi-elastic signal at $\boldsymbol{Q}_{\mathrm{SDW}}$ in superconducting regions confirms the absence of the SDW in the superconducting phase. In contrast, the oxygen deficient sample develops both a dispersive magnon and a clear enhancement of quasi-elastic intensity at $\boldsymbol{Q}_{\mathrm{SDW}}$ that resembles the bulk ambient pressure phase (**Supplementary Fig. 5, 6**). The spatial segregation between superconductivity and SDW order, together with the absence of a detectable static SDW and associated magnetic excitations in superconducting regions, indicates that SDW order is extrinsic to superconductivity in LPNO. Superconductivity therefore does not emerge from a SDW parent state. Instead, oxygen deficiency stabilizes the (0.25, 0.25) SDW order, at least in the case of compressively strained thin films.

**Distinct electronic ground states via oxygen stoichiometry**

Having established that superconductivity in LPNO is associated with oxygen-stoichiometric, SDW-free regions, we probe this state using Ni-$L_3$ and O-$K$ edge x-ray absorption spectroscopy (XAS) and resonant inelastic x-ray scattering (RIXS), which provide elemental-specific sensitivity to Ni and O orbital character. **Figure 3a** shows polarization-resolved Ni-$L_3$ XAS for superconducting and O-def films after subtracting the strong La $M_4$ contribution (See **Supplementary Fig. 7**). All spectra from the superconducting film were acquired in SDW-free regions (near spot A in **Fig. 1c**). At an incident angle of $\theta$ = 10°, x-rays with σ/π polarization primarily probe in-plane (ε//$ab$)/out-of-plane (ε//$c$) orbitals, respectively. For ε//$c$, the SC film exhibits a dominant peak at 853 eV followed by an extended high-energy tail, whereas the O-def film develops an additional shoulder near 853.5 eV. For ε//$ab$, the leading peak in XAS is at slightly higher energy (853.23 eV), above which a broadened high-energy feature around 854-855 eV is observed that is strongly suppressed in the O-def film. This high-energy spectral weight is commonly associated with enhanced Ni-O covalency and itinerant character – it is enhanced in superconducting films with low normal state resistivity[26], whereas its suppression often corresponds to less metallic or insulating samples[16,31].

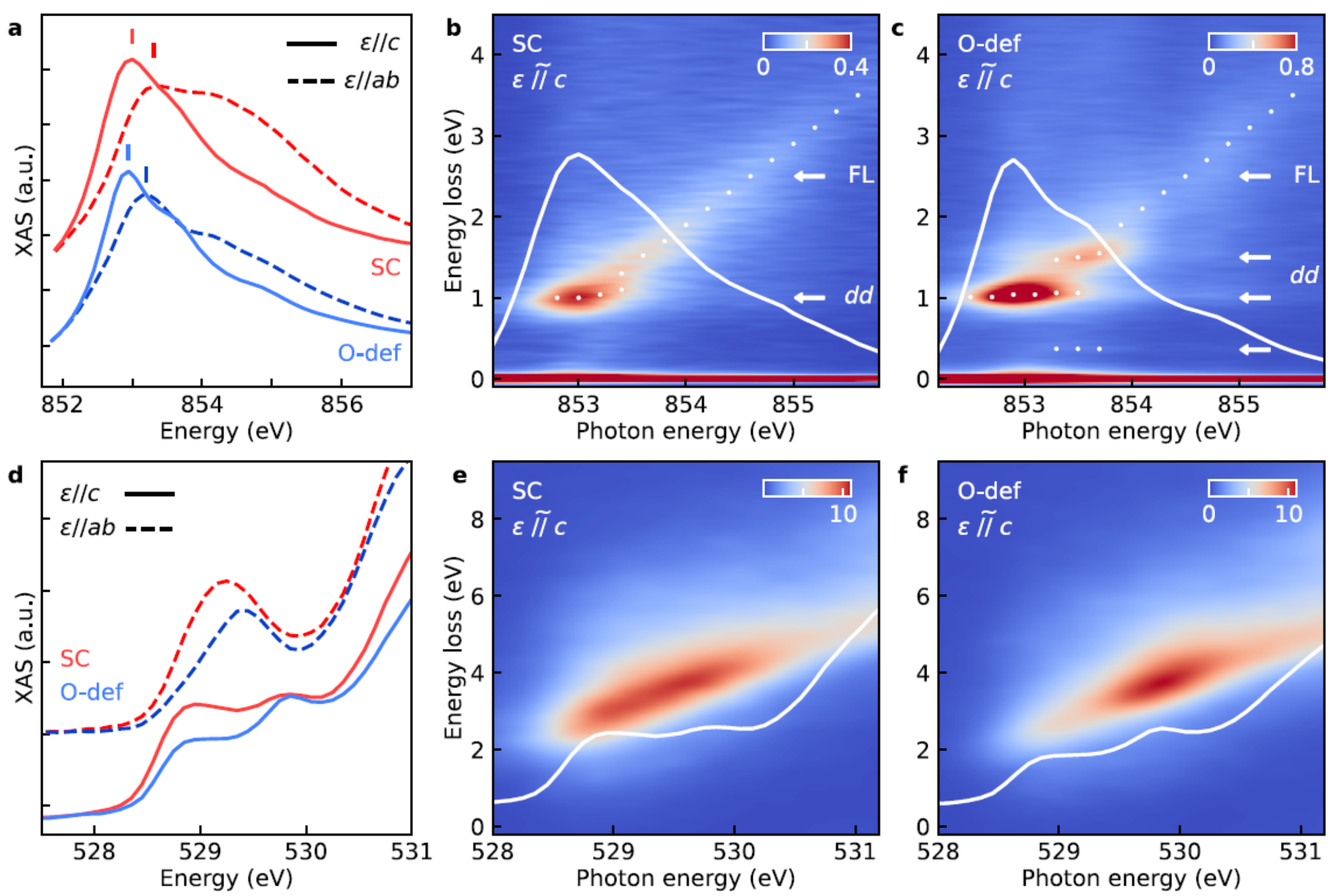

**Figure 3. X-ray absorption spectra (XAS) and out-of-plane resonant inelastic x-ray scattering spectra (RIXS) of LPNO at Ni-$L_3$ and O-$K$ edges. a**, Polarization-dependent Ni-$L_3$ XAS after subtracting the La $M_4$ contribution (**Supplementary Fig. 4**). Vertical ticks mark the leading XAS maxima in both polarization channels. Traces are offset for clarity. **b, c**, Incident-energy dependent RIXS maps for SC (**b**) and O-def (**c**) films with polarization nearly along the *c*-axis of the samples (ε//*c*). The major mode positions are marked by white dots. The corresponding XAS are superimposed in white. **d,** Polarization-dependent O-*K* XAS near the pre-peak region for superconducting and oxygen deficient LPNO. **e, f,** Out-of-plane incident energy RIXS maps near O-*K* edge. The corresponding XAS are superimposed in white.

We further resolve the oxygen-stoichiometry-driven changes in Ni local electronic structure using Ni-$L_3$ RIXS. In the superconducting film, a Raman-like *dd* excitation appears at 1 eV energy loss (**Fig. 3b**), which mimics the spectroscopic signature of a Ni $3d^8$ configuration and corresponds to localized crystal-field excitations between Ni $t_{2g}$ to $e_g$ manifolds, consistent with the energy scale observed in other nickelate families[33–35]. The *dd* excitation is followed by a prominent fluorescence-like continuum at higher energy loss that tracks incident energy and resonates on the high-energy component in the XAS (853.6 eV), indicating substantial weight from delocalized, strongly hybridized Ni-O states. In the O-def film, the *dd* excitations become substantially more intense in both polarization channels (**Fig. 3c** and **Supplementary Fig. 8**), with two additional modes emerging at 0.4 eV and 1.5 eV in the ε//*c* channel, indicating a more localized Ni electronic configuration along the *c*-axis. A direct comparison of ε//*c* spectra at selected incident energies highlights the oxygen-stoichiometry-driven changes in RIXS spectra (**Supplementary Fig. 8e**), where the *dd* excitations are sharper and nearly three times more pronounced, spanning a wider resonance range. Furthermore, while the *dd* excitation at 1.5 eV is Raman-like in the O-def film, this feature loses its Raman character and becomes fluorescent in the superconducting film. Together, the Ni-edge results indicate that oxygen deficiency pushes LPNO towards a more localized, less ligand-hybridized electronic structure, whereas superconducting films remain in a more itinerant, strongly hybridized regime. Previously reported Ni-$L_3$ RIXS spectra from bulk and some thin film samples resemble our O-def film[14,30,32], which may have a similar origin.

Within the Zaanen-Sawatzky-Allen (ZSA) framework, the low-energy electronic structure of late transition-metal oxides is governed by the competition between the on-site Coulomb repulsion U and Δ, the charge-transfer energy[36]. The nominal $Ni^{2.5+}$ valence positions bilayer nickelates in between $LaNiO_3$ (nominal $Ni^{3+}$) and NiO or $La_2NiO_4$ ($Ni^{2+}$). Strong covalency can push the system into the small- or negative-charge-transfer regime ($\Delta \sim 0$), with ligand-hole configuration $d^8\underline{L}$ acquiring substantial weight in the ground state. O-*K* edge XAS and RIXS directly probe unoccupied O 2*p* states from the transition between O 1*s* core level and 2*p* states hybridized with Ni 3*d* and thus provide a direct view of ligand-hole character across oxygen stoichiometry (**Fig. 3d-f** and **Supplementary Fig. 9**). Both films show a clear pre-edge feature in XAS, indicating a substantial oxygen character in the ground state wavefunction of LPNO. In the ε//*ab* channel, which probes in-plane O $2p_{x,y}$ hybridized with Ni $3d_{x2-y2}$, the SC film displays a well-defined pre-edge peak at 529.2 eV. With oxygen deficiency, this pre-edge peak decreases in intensity and shifts to higher energy. The oxygen stoichiometry-dependence of the ε//*ab* pre-edge peak is reminiscent of the doping-dependence of the Zhang-Rice singlet state in cuprates that holds strong charge transfer character[37,38]. The unoccupied projected density of states in the in-plane O $2p_{x,y}$ orbitals is estimated around 10%, based on the doping-dependent pre-edge peak in doped cuprates[39]. The ε//*ab* RIXS map, which probes predominantly in-plane orbitals, shows a prominent charge-transfer excitation continuum spanning 2.5 ~ 5 eV energy loss (**Supplementary Fig. 9**) that mirrors $LaNiO_3$ (**Supplementary Fig. 10**), underscoring a highly similar electronic configuration governed by small or negative charge-transfer energy[27,31].

In the ε//*c* channel, two clear pre-edge peaks at the O-*K* edge indicate hole character associated with O $2p_z$ character. The intensity of the lower-energy pre-edge peak is sensitive to oxygen deficiency without an energy shift, while the high energy pre-peak is less sensitive to oxygen deficiency. The ε//*c* RIXS map further probes the oxygen-stoichiometry-induced changes in electronic structure, where a strong, fluorescence-like emission is observed at 2 ~ 6 eV energy loss (**Fig. 3e, f**), highlighting the covalent nature of the O $2p_z$ state. Compared to the O-def sample, the intensity of the leading excitation in the ε//*c* is enhanced in the superconducting sample, suggesting the presence of higher ligand hole density along the *c*-axis. Since the outer apical oxygen does not carry a ligand hole[27], this enhanced hole density must be primarily associated with the inner apical O $2p_z$ ligand, which bridges two Ni $3d_{z2}$ orbitals.

**Apical oxygen hole stabilized interlayer five-spin polaron**

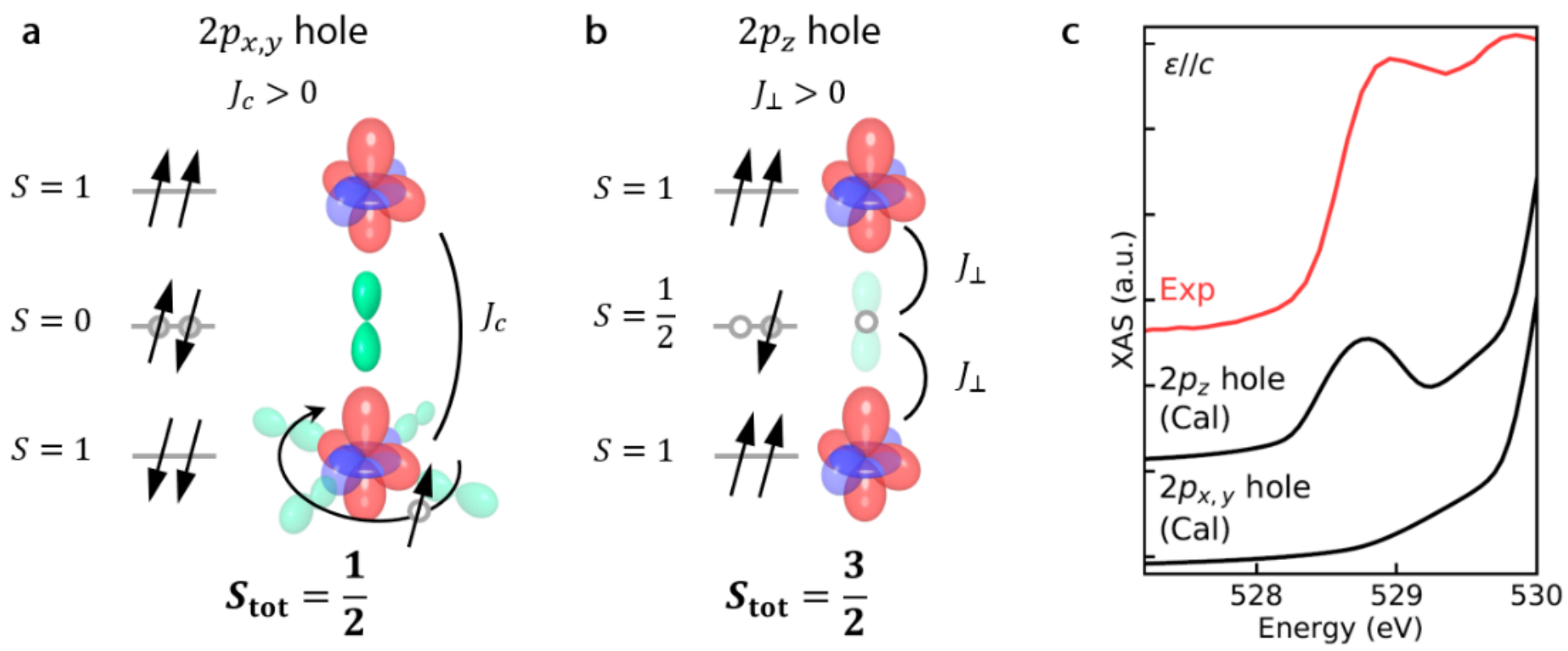

**Figure 4. Possible electronic and magnetic structure of the Ni dimer in stoichiometric LPNO containing 5 holes. a**, A ligand hole occupying $2p_{x,y}$ orbitals, leading to a $S_{tot} = 1/2$ ground state. **b**, A ligand hole occupying $2p_z$ orbital of apical oxygen, forming a 5-spin polaron ground state with $S_{tot} = 3/2$. **c**, Comparison of experimental (red) and calculated O-*K* XAS (black) based on the configurations in **a** and **b**. Only the 5-spin polaron configuration with $S_{tot} = 3/2$ reproduces the O-*K* pre-edge features in the experiment.

In stoichiometric bilayer nickelates, each bilayer structural unit hosts nominally five holes shared between two Ni sites. For Ni with valence higher than 2+, it typically takes two holes to form $3d^8$ configuration, with the rest of the holes going to oxygen ligand orbitals[33]. The distribution of the last remaining hole is an important question, as it is intimately related to the electronic and magnetic ground state of bilayer nickelates. **Figure 4** shows two simplified scenarios where the ligand hole occupies either the planar oxygen $2p_{x,y}$ orbitals or the apical $2p_z$ orbital. For planar $2p_{x,y}$ holes (no apical holes, **Fig. 4a**), a singlet ground state is expected whereby the two Ni atoms consist of localized spin-1 states coupled antiferromagnetically via virtual $3d_{z2}$-$2p_z$-$3d_{z2}$ interlayer exchange interaction ($J_c$) across the apical oxygen. Meanwhile, the planar dynamics is characterized by strongly hybridized $3d_{x2-y2}$-$2p_{x,y}$ hole states. In the other scenario, as shown in **Fig. 4b**, the hole resides in the $2p_z$ orbital of the connecting apical oxygen. The exchange interaction with apical ligand hole ($J_\perp$) aligns the two $S = 1$ Ni spins to be parallel, forming a coupled interlayer quartet 5-spin polaron with total spin $S_{tot} = 3/2$ (see **Methods**).

While some ligand hole weight is present in the planar $2p_{x,y}$, the O-*K* XAS pre-edge spectral weight in ε//*ab* channel indicates around 10% ligand hole. Therefore, the planar $2p_{x,y}$ holes scenario cannot fully account for the experiment data, and most of the hole wavefunction must reside in the bridging apical oxygen. Indeed, our ε//*c* O-*K* edge XAS and RIXS results (**Fig. 3**) suggest the presence of a significant hole density in the apical $2p_z$ orbital for the superconducting sample (i.e. stoichiometric compound), indicating the relevance of the 5-spin polaron scenario.

To validate this picture, we model the XAS and RIXS for both scenarios using the Bethe–Salpeter equation (BSE)-based OCEAN code for core-level Oxygen $K$-edge spectroscopy[40]. Initial ground-state wavefunctions are generated from the Quantum ESPRESSO density functional theory (DFT) package with SCAN exchange–correlation functional[41]. The two $S = 1$ Ni spins are forced to be antiparallel (parallel) for the $2p_{x,y}$ ($2p_z$) hole configurations. The calculated $\varepsilon//c$ O-$K$ XAS are shown in **Fig. 4c**. With the apical $2p_z$ hole, the $\varepsilon//c$ XAS shows a strong O-$K$ pre-edge peak at 528.9 eV, consistent with the experiment, whereas the $2p_{x,y}$ hole does not produce the same pre-edge feature. The calculated RIXS spectra for the 5-spin polaron state also reproduces the experimental RIXS spectra (**Supplementary Fig. 11**).

The planar versus axial polarization-dependent XAS in the SC and O-def samples may be understood in this context. The shift to lower energies of the in-plane XAS in the SC spectra compared to the deficient one indicates that some holes are doped into the Ni-O plane, reminiscent of the Zhang-Rice singlet state in the cuprates. On the other hand, the similarity of the peak positions of the axial polarization dependent XAS, along with the overall growth of the signal in SC versus oxygen deficient spectra, indicate that holes are always present in the bridging oxygen ligand orbitals. The overall signal grows accordingly with the number of bridging apical sites.

The robustness of this interlayer 5-spin polaron state is determined by the coupling strength $J_\perp$, which can be estimated from the excitation energies within the local 5-spin state. This 5-spin polaron is more strongly coupled compared to the AF singlet configuration due to the exchange between neighboring Ni-apical spins, $J_c/J_\perp \sim (t_\perp/\Delta)^2$, with $t_\perp$ the $2p_z$-$3d_{z2}$ hybridization and $\Delta$ the axial charge transfer energy. Our Ni-$L_3$ RIXS data for the superconducting sample shows no orbital excitations below the 1 eV *dd* excitation, suggesting that the spin excitations likely overlap with and is obscured by the *dd* excitation manifold. We note that the enhanced fluorescence feature in the Ni-$L_3$ RIXS map with $\varepsilon//c$ in the energy range of 1 ~ 2 eV (**Fig. 3b**) may be a signature of these excitations, which establishes a lower bound for $J_\perp$, indicating that the interlayer 5-spin polaron state is a robust bound state. As a result, holes fill the bridging apical site first (if present), locking the electronic and spin states of the $3d_{z2}$ orbitals, leaving the half-filled $d_{x2-y2}$ orbital plus remaining ligand holes to dictate the low-energy physics—a scenario reminiscent of cuprates and infinite-layer nickelates[1,42].

Taken together, our results demonstrate the sensitivity of the electronic and magnetic ground state in bilayer nickelates to oxygen stoichiometry and ligand hole distribution. We identify oxygen deficiency as the primary driving force for the SDW phase in bilayer nickelates, which is spatially segregated from the oxygen-stoichiometric, SDW-free superconducting phase. Furthermore, $\varepsilon//c$ x-ray spectroscopy reveals substantial ligand hole occupancy at the apical oxygen site, pointing to an interlayer 5-spin polaron ground state in superconducting LPNO. This observation constrains the possibilities by which SDW magnetism plays an essential role for superconductivity, and instead suggests the apical-ligand-mediated 5-spin polaron as a key feature of superconducting bilayer nickelates. These insights establish the precision control of out-of-plane correlations as a critical parameter for the design of superconductivity in multilayer Ruddlesden–Popper nickelate superconductors.

Upon completion of this work, we became aware of other related studies[43,44].

## Methods:

### Sample preparation

The $La_2PrNi_2O_{7-x}$ (~5 nm) thin films were synthesized on (001)-oriented $SrLaAlO_4$ (SLAO) substrates via pulsed laser deposition using a KrF excimer laser (248 nm wavelength). Post-growth ozone annealing was carried out to achieve optimal oxygen stoichiometry in the sample. Details of the sample preparation are provided in Ref. 23. Synchrotron x-ray diffraction was carried out at Stanford Synchrotron Radiation Lightsource (SSRL) Beamline 17-2 and confirmed the films were coherently strained to the underlying substrates (**Supplementary Fig. 12**).

To prevent degradation of superconductivity due to oxygen loss under ambient conditions, all films were maintained under cryogenic conditions (77 K) for storage and transported in a dry-ice environment (194 K). The quality of the superconducting state and oxygen content were verified by transport measurements conducted immediately before and after the x-ray experiments. $T_c$ and the normal-state resistivity of each film were monitored before and after the x-ray measurements (**Supplementary Fig. 13**), confirming that oxygen stoichiometry, and thus the oxygen-stoichiometric superconducting state, remain stable throughout the reported measurements.

### Resonant x-ray scattering

Resonant x-ray scattering measurements were conducted at SSRL Beamline 13-3 and the REIXS (10ID-2) beamline at the Canadian Light Source (CLS), with the base temperature of 11 K, well below the SDW transition temperature. The incident photon energy was tuned to the Ni-$L_3$ resonance ~853 eV with π polarization to maximize the SDW signal. The scattered x-rays were collected at a scattering angle of 150° using a 2D CCD detector (SSRL) or microchannel plate (CLS). The raw scattering intensity was converted to reciprocal space coordinates using the known in-plane lattice constants of the substrates (SLAO: $a = b = 3.756$ Å).

Spatial maps of the SDW intensity were acquired by raster-scanning a ~200 μm x-ray beam across the sample surface. During the scan, a fixed scattering geometry was maintained to probe the (0.25, 0.25) SDW peak. For each point in the spatial map, the local SDW intensity was obtained by fitting the reciprocal space intensity with a 2D Gaussian peak profile and a constant background to account for isotropic fluorescence.

### High-resolution resonant inelastic x-ray scattering (RIXS)

X-ray absorption spectra (XAS) and resonant inelastic x-ray scattering (RIXS) experiments were performed at the SIX (2-ID) of the National Synchrotron Light Source II (NSLS-II). XAS data were taken in total fluorescence yield mode with x-ray incident at a grazing angle of 10°. RIXS spectra were collected using a fixed spectrometer angle $2\theta = 150°$, with the combined energy resolution optimized at around 24 meV at Ni-$L_3$ edge. Part of the O-*K* RIXS spectra were collected at the iRIXS endstation at Beamline 8.0.1 of the Advanced Light Source (ALS) at Lawrence Berkeley National Laboratory using a fixed spectrometer angle $2\theta = 90°$, $\pi$ polarization and temperature of 80 K, with the energy resolution around 0.2 eV at O-*K* edge.

**BSE-based OCEAN and Quantum ESPRESSO DFT calculations**

The OCEAN (v. 3.2.1) code, which uses the BSE for many-body effects in spectroscopy, was used to calculate the oxygen-*K* edge transitions in LPNO[40,45]. Input wavefunctions were generated using the Quantum ESPRESSO (v. 7.4) DFT package. We used the meta-generalized gradient approximation SCAN exchange–correlation functional[46]. We generated pseudopotentials via the scalar-relativistic optimized norm-conserving Vanderbilt pseudopotentials code (ONCVPSP v. 3.3.1) within the projector-augmented wave and Perdew–Burke–Ernzerhof formalism[47]. Convergence criterion for the electronic self-consistent field calculation was $10^{-10}$ Ry and for the plane-wave cut-off, we used components up to 110 Ry. A small total energy difference $\Delta E_{\text{AF-F}}$ = -23 meV/atom was obtained for the axial AF aligned Ni configuration and ferromagnetic aligned configuration. A gamma-centered *k*-grid of 6 × 6 × 2 was used, and incoming and outgoing photon polarization were considered to match the experimental configuration. A global energy shift was also added to all the calculated spectral features for ease of comparison with experiment.

**Spin model for five-spin polaron**

We consider a distribution of five holes across two Ni sites bridged by an apical oxygen along the *z*-axis. The relevant orbitals are the Ni $e_g$ and apical $p_z$ orbitals. For a scenario where the last hole resides in the apical $p_z$, the effective low energy spin Hamiltonian is given by (T, B denotes top and bottom Ni spins, respectively, and $p$ denotes the bridging apical spin):

$$\widehat{H}_{\text{eff}} = \Delta + J_{\perp}\left[\left(\widehat{\boldsymbol{S}}_{\text{T},z^2} + \widehat{\boldsymbol{S}}_{\text{B},z^2}\right)\cdot\widehat{\boldsymbol{S}}_p - \frac{1}{2}\right] - J_{\text{H}}\left[\widehat{\boldsymbol{S}}_{\text{T},x^2-y^2}\cdot\widehat{\boldsymbol{S}}_{\text{T},z^2} + \widehat{\boldsymbol{S}}_{\text{B},x^2-y^2}\cdot\widehat{\boldsymbol{S}}_{\text{B},z^2}\right]$$

with $\Delta$ the oxygen site energy relative to Ni $e_g$ orbitals, $J_{\text{H}}$ the intra-atomic Hund's exchange, and $J_{\perp}$ the exchange coupling between the Ni spins and the oxygen spin

$$J_{\perp} = 2t_{\perp}^2\left[\frac{1}{2U_{dd} - 2J_{\text{H}} - \Delta} + \frac{1}{U_{pp} + \Delta - U_{dd} + 3J_{\text{H}}}\right]$$

where $t_{\perp}$ is the Ni-O hopping. $J_{\perp}$ tends to align the Ni $z^{\mathbf{2}}$ spins antiferromagnetically with the ligand spin to make an interlayer high spin polaron, while $J_{\text{H}}$ favors ferromagnetic alignment of the Ni $z^2$ *and* $x^2 - y^2$ spins. Assuming Hund's energy is large enough to lock each Ni in the $S$ = 1 configuration, the spin Hamiltonian can be rewritten as

$$\hat{H}_{\text{eff}} = \Delta - \frac{J_{\text{H}}}{2} + J_{\perp}\left[\hat{\boldsymbol{S}}_d^{\text{T}} \cdot \hat{\boldsymbol{S}}_p - \frac{1}{2}\right]$$

with $\hat{\boldsymbol{S}}_d^{\text{T}} = \hat{\boldsymbol{S}}_{\text{T}} + \hat{\boldsymbol{S}}_{\text{B}}$ the total spin operator formed by two Ni spin 1 operators ($\hat{\boldsymbol{S}}_{\text{T}}$, $\hat{\boldsymbol{S}}_{\text{B}}$), with spin quantum number $S_d = 2, 1, 0$. The total spin quantum number can take on the values $S_{\text{tot}} = \frac{5}{2}, \frac{3}{2}, \frac{1}{2}$. Thus, the eigenvalues are

$$\text{Sextet } \left(S_{\text{tot}} = \frac{5}{2}, S_d = 2\right): E_{\text{S}} = \Delta - \frac{J_{\text{H}}}{2}$$

$$\text{Quartet } \left(S_{\text{tot}} = \frac{3}{2}, S_d = 2\right): E_{\text{Q}} = \Delta - \frac{J_{\text{H}}}{2} - \frac{5J_{\perp}}{4}$$

$$\text{Quartet } \left(S_{\text{tot}} = \frac{3}{2}, S_d = 1\right): E_{\text{Q}} = \Delta - \frac{J_{\text{H}}}{2} - \frac{J_{\perp}}{4}$$

$$\text{Doublet } \left(S_{\text{tot}} = \frac{1}{2}, S_d = 1\right): E_{\text{D}} = \Delta - \frac{J_{\text{H}}}{2} - J_{\perp}$$

$$\text{Doublet } \left(S_{\text{tot}} = \frac{1}{2}, S_d = 0\right): E_{\text{D}} = \Delta - \frac{J_{\text{H}}}{2} - \frac{J_{\perp}}{2}$$

Thus, the ground state is a four-fold degenerate quartet, effectively forming a 5-spin polaron via oxygen-mediated ferromagnetically aligned Ni spins, stabilized by Hund's coupling. The excitations probe-able via RIXS are $\Delta E = \frac{J_{\perp}}{4}, \frac{3J_{\perp}}{4}, \frac{5J_{\perp}}{4}$ with $\Delta S = 1$ (spin flip), and $\Delta E = J_{\perp}$ for $\Delta S = 0$ (non-spin flip).

**Spin model for in-plane hole configuration**

If we instead examine a 4-hole local configuration, whereby the 5th hole resides in the Ni-O *ab* plane, an effective low energy Hamiltonian can be rewritten as

$$\hat{H}_{\text{eff}} = -\frac{J_{\text{H}}}{2} + J_{\perp}^{\text{eff}}\left(\hat{\boldsymbol{S}}_{\text{T},z^2} \cdot \hat{\boldsymbol{S}}_{\text{B},z^2} - \frac{1}{4}\right)$$

with

$$J_{\perp}^{\text{eff}} = \frac{4t_{\perp}^4}{\Delta^2}\left(\frac{1}{U_{dd}} + \frac{2}{2\Delta + U_{pp}}\right).$$

Similarly, the eigenvalues are

$$\text{Quintet } (S_{\text{tot}} = 2): E_{\text{Q}} = -\frac{J_{\text{H}}}{2}$$

$$\text{Triplet } (S_{\text{tot}} = 1): E_{\text{T}} = -\frac{J_{\text{H}}}{2} - \frac{J_{\perp}^{\text{eff}}}{2}$$

$$\text{Singlet } (S_{\mathrm{tot}} = 0): E_{\mathrm{S}} = -\frac{J_{\mathrm{H}}}{2} - \frac{3J_{\perp}^{\mathrm{eff}}}{4}$$

The ground state for the 4-hole configuration is an inter-Ni spin singlet made from two $S = 1$ Ni moments. Therefore, the total spin is $S_{\mathrm{tot}} = \frac{1}{2}$ after incorporating the 5th hole in the planar O orbitals. This is qualitatively different from the 5-spin polaron case. The excitations are $\Delta E = \frac{J_{\perp}^{\mathrm{eff}}}{4}$ for $\Delta S = 1$ and $\Delta E = \frac{3J_{\perp}^{\mathrm{eff}}}{4}$ for $\Delta S = 2$.

**Acknowledgements:**

We thank S. Kivelson, S. Raghu, Y. Wu, H. Oh, B. Wang, Z.X. Shen, E. Zhang for fruitful discussions. This work was supported by the US Department of Energy (DOE), Office of Science, Basic Energy Sciences, Materials Sciences and Engineering Division, under contract no. DE-AC02-76SF00515. This research used beamline 2-ID of NSLS-II, a US DOE Office of Science User Facility operated for the DOE Office of Science by Brookhaven National Laboratory under Contract No. DE-SC0012704. Part of x-ray measurements were carried out at beamlines 13-3 and 17-2 of SSRL, SLAC National Accelerator Laboratory, supported by the US Department of Energy, Office of Science, Office of Basic Energy Sciences (contract no. DE-AC02-76SF00515). Part of the research was conducted at the Advanced Light Source (ALS), which is a DOE Office of Science User Facility, under contract no. DE-AC0205CH11231. Part of the research described in this paper was performed at the Canadian Light Source, a national research facility of the University of Saskatchewan, which is supported by the Canada Foundation for Innovation (CFI), NSERC, the National Research Council (NRC), the Canadian Institutes of Health Research (CIHR), the Government of Saskatchewan, and the University of Saskatchewan. Calculations were performed on resources of the Sherlock cluster at Stanford University, and on resources of the National Energy Research Scientific Computing Center (NERSC), a Department of Energy Office of Science User Facility, using NERSC award BES-ERCAP0031424.